\documentclass[twocolumn,pra]{revtex4}

\usepackage{graphicx}
\usepackage{dcolumn}
\usepackage{bm}

\newcommand {\deriv}[3]{\frac{\partial^#1#2}{\partial #3^#1}}
\newcommand{\etal}{{\it et al\ }}

\def \del {\delta}
\def \thet {\theta}
\def\mO{\mbox{O}}
\def\smO{\mbox{{\tiny O}}}

\begin{document}

\title{Calculation of muon transfer from muonic hydrogen to atomic oxygen} 

\author{Arnaud Dupays, Bruno Lepetit,
J.\ Alberto\ Beswick, Carlo Rizzo}
\affiliation{
Laboratoire Collisions, Agrégats, Réactivité, IRSAMC,
Universit\'e P. Sabatier, 31062 Toulouse, France
}%

\author{
Dimitar Bakalov}
\affiliation{%
INRNE, Bulgarian Academy of Sciences, Sofia, Bulgaria
}%

\date{\today}

\begin{abstract}
The muon transfer probabilities  between muonic
hydrogen and an oxygen atom 
are calculated in a constrained geometry one dimensional model
for collision  energies between $10^{-6}$ and $10^3$ eV.
For relative translational energies below $10^{-1}$ eV, for which the de Broglie wavelength 
($>1$ \AA) is much larger
than the characteristic distance of the potential interaction ($\sim 0.1$ \AA),
the problem corresponds
to an ultra-cold collision.
The close-coupling time-independent quantum equations
are written in terms of hyperspherical coordinates and 
a diabatic-by-sectors basis set. 
The muon transfer probabilities are qualitatively interpreted in terms of a model involving two Landau-Zener crossings  
together with the threshold energy dependence.
Based on this analysis a simple procedure to estimate the energy dependence
 of the muon transfer rate in three dimensions, is proposed.
These estimated rates are discussed in the light of previous model calculations
 and available experimental data for this process.
 It is concluded that
the high transfer rates at epithermal energies inferred from experiments
are unlikely to be correct.
\end{abstract}
\maketitle

\section{Introduction} 

Negative muon transfer between exotic atoms (muonic hydrogen, for instance) 
and other atoms or molecules, has been extensively studied in the framework
of muon catalysed nuclear fusion (see Ref.~\cite{Pono:01} and literature cited therein). Also, the structural and spectroscopic
properties of these species
 is of interest for metrology as well
as a test of quantum electrodynamic theories \cite{Baka:93}.

Very recently,  a new method  to measure the hyperfine structure  ($F=0,1$) of
muonic hydrogen ($p\,\mu$) based on the collisional energy dependence of
muon transfer from the muonic hydrogen to an oxygen molecule,
has been proposed \cite{Adam:01}.
When muonic hydrogen in the $F=0$ hyperfine ground state is laser excited to
the $F=1$  level, collisions with $H_2$ convert this
excess energy  into kinetic energy, giving an additional
0.12 eV  translational energy to the muonic hydrogen. When
the muon is transfered to an oxygen atom, it is captured in
high ($n=5, 6$) states which promptly de-excite and emit X-rays.
If the muon transfer rate
to oxygen changes significantly from thermal (0.04 eV) to epithermal
(0.16 eV) energies, the measurement of the X-ray emission intensity with and without
laser excitation can be used for the determination of the hyperfine splitting
in muonic hydrogen \cite{Baka:93,Adam:01}.
This proposal was based on the  work of Werthm\"uller \etal \cite{Wert:98}  which indicates that  muon transfer 
from hydrogen to oxygen  increases by  a factor of 3-4  going from thermal
 to epithermal  collision energies. 

While the experimental transfer rate at thermal energies have been borne out by the calculations
of Sultanov and Adhikari \cite{Sult:00}, the rates at epithermal energies have never been confirmed 
theoretically. Even more, such an increase of the transfer rate with energy does not have
a clear theoretical explanation. Werthm\"uller \etal  \cite{Wert:98} suggest
the existence of a resonance at low energies. It is worth to stress that such an increase
has not been observed in the case of muonic hydrogen colliding with sulfur \cite{Mulh:93}
while for the case of CH$_4$, experimental data suggest that the
transfer rate actually decreases going from thermal to epithermal energies \cite{Kirc:98}.
In order to improve our understanding of this process and   to assess the validity 
of the proposal of  Ref.~\cite{Adam:01}, it is important to perform model  calculations to
study in detail this reaction as a function of energy. 
It is the purpose of this
paper to  present such a study.

Since
the muonic hydrogen has to approach one of the oxygen  nuclei
very close in order for the muon to transfer  \cite{Gers:63}, 
 the process can be described as
\begin{equation}\label{I1}
(p\mu)+\mO^{8+}\to p+(\mu\mO)^{7+}
\end{equation}
 
Although there have been several full three-dimensional calculations
of muon transfer rates at low energies between muonic-hydrogen and low-$Z$ atoms
(see literature cited in Ref.~\cite{Sult:02}),
there is none when the transfer involves nuclei with $Z>3$ . There are several good reasons
for that.  As  $Z$
 increases ($Z=8$ in our case) there is a larger initial-channel polarization
and a stronger final-channels Coulomb interaction. In addition the number of open channels
even at zero relative kinetic energy increases with $Z$ making the full three-dimensional
calculation very heavy. Thus up-to-now only approximate calculations have
been performed for the muon transfer rate between muonic-hydrogen and 
oxygen \cite{Haff:77,Gers:63,Sult:00}.  Among them, the most recent and accurate is the one 
done by Sultanov and Adhikari \cite{Sult:00}, who used a two-channel
approximation to the integro-differential Fadeev-Hahn formalism to calculate the muon transfer rates
from muonic-hydrogen to the $n=5$ states of oxygen.

We have chosen to perform exact calculations but 
in a restricted 
configuration with the muon moving along the line joining the
proton to the oxygen atom.  
Indeed, for this system the colinear configuration
is the most favorable for the transfer process. Furthermore,
since  the colinear configuration provides
the minimum energy path for the reactance channel, even for an initial non colinear configuration
there will
be efficient orientational effects  in particular for low translational energies.


Two model interaction potentials have been used in the calculations:
\begin{enumerate}
\item A pure coulombic potential  with the 
 bare oxygen nucleus $\mO^{8+}$. 

\item A shorter range model potential  with a distance dependent effective charge on the oxygen
nucleus
(calculated via a Thomas-Fermi electronic density model) 
to simulate the effect of the outer electrons.

\end{enumerate}

The Hamiltonian
was written in terms of hyperspherical coordinates.
A piecewise diabatic basis set on the hyperspherical angle
 was used to expand the wave function.
The basis functions have in turn been expanded in terms of first order Legendre functions
 to efficiently handle
 the two coulombic singularities at fixed hyperradii.
The resulting close-coupling time-independent
Schr\"odinger equations in the hyperradius were solved using a de Vogelaere algorithm
and the partial and total muon transfer probabilities were determined by the standard
S-matrix analysis at large distances.
Since for energies below $10^{-1}$ eV the muon transfer process studied here is equivalent
to an ultra-cold collision (de Broglie wavelength ($>1$ \AA) much larger
than the characteristic distance ($\sim 0.1$ \AA) of the potential interaction), special care had to be
taken to the asymptotic analysis in the reactant channel.

The paper is organized as follows. Section 2 introduces the model and the methodology
used in the calculations. Section 3 presents the calculated muon transfer probabilities
together with 
their  interpretation in terms
of simple Landau-Zener and threshold models. A  procedure to estimate their energy behavior
in the 3-dimensional case,
is also presented. In section 4, the muon transfer rates are 
discussed in the light of previous model calculations
 and available experimental data for this process.
Finally, section 5 is devoted to the conclusions.


\section{Methodology}

 The mass-scaled Jacobi coordinates
adapted to the entrance channel of reaction (\ref{I1}) are~:

\begin{eqnarray}\label{M1}
R &=& \sqrt{\frac{m_{\smO,p\mu}}{m}}\,\left(x_{\smO} - \frac{m_\mu\,x_\mu + m_p\,x_p}{m_p+m_\mu}\right);
\\ \label{M1b}
r &=& \sqrt{\frac{m_{p,\mu}}{m}}\,(x_\mu - x_p)
\end{eqnarray}
where

\begin{eqnarray}\label{M2}
m_{\smO,p\mu} &=& \frac{m_{\smO}\,(m_p + m_\mu)}{m_{\smO} + m_p + m_\mu};
\quad m_{p,\mu} =  \frac{m_p\,m_\mu}{ m_p + m_\mu};
\\ \label{M2b}
 m &=& \left(
\frac{m_{\smO}\,m_p\,m_\mu}{m_{\smO} + m_p + m_\mu}
\right)^{1/2}
\end{eqnarray}

The hyperspherical coordinates are then defined by

\begin{equation}\label{M3}
\rho = \sqrt{R^2 + r^2};\quad \tan\,\thet = r/R
\end{equation}
where $0\le \theta\le \theta_{\mu}$ , with $\theta_{\mu}=\mbox{arctan}(m_{\mu}/m)$. In terms of these
coordinates and after regularisation of the wave function by the factor $\sqrt{\rho}$ (the volume element is then
given by $d\rho\,d\thet$), the time-independent Schr\"odinger equation at total energy 
$E$ is:

\begin{eqnarray}
-\frac{\hbar^2}{2m}\,\left(
\deriv2{}{\rho} + \frac{1}{\rho^2}\,\deriv2{}{\thet}\right)\,\psi(\rho,\thet)
\nonumber \\
+ 
\left(
 V(\rho,\thet) - \frac{\hbar^2}{8m\rho^2} - E \right)\,\psi(\rho,\thet)  = 0
\label{M5} 
\end{eqnarray}
where $V(\rho,\thet)$ is the interaction potential given by

\begin{eqnarray}
V = -\frac{e^2\,Z_{\smO}^\ast(\vert x_\mu - x_{\smO}\vert)}{\vert x_\mu - x_{\smO}\vert} \nonumber \\
+ 
\frac{e^2\,Z_{\smO}^\ast(\vert x_p - x_{\smO}\vert)}{\vert x_p - x_{\smO}\vert} 
- \frac{e^2}{\vert x_p - x_\mu\vert}
\label{M6}
\end{eqnarray}

\begin{figure}
\includegraphics[width=8.cm]{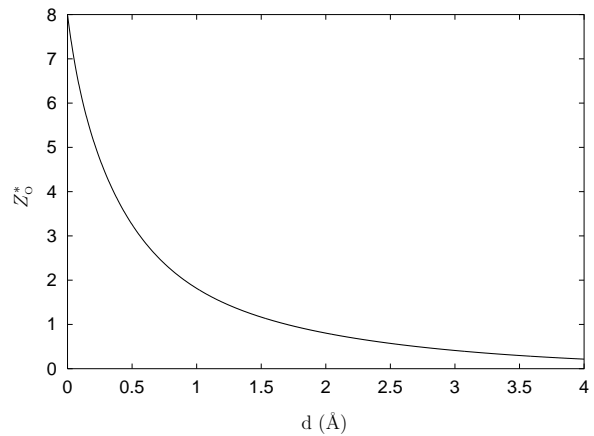}
\caption{\label{figure_a1_1}{Thomas-Fermi effective charge for the oxygen atom as a function of the distance
to the O$^{8+}$ nuclei.}}
\end{figure}

Two sets of calculations have been performed. One with the bare oxygen nucleus,
for which $Z_{\smO}^\ast=8$ in (\ref{M6}), and another
with $Z_{\smO}^\ast(d)$ being an effective charge given by the Thomas-Fermi model \cite{Land:77}
and represented in Fig.~\ref{figure_a1_1}.

The total wave function $\psi(\rho,\thet)$ is expanded in terms of 
basis set wave functions depending on the hyperspherical angle
$\thet$.
We use a diabatic-by-sector
representation. In each sector $\rho_n-\del\rho_n \leq \rho < \rho_n+\del\rho_n; n=1,...,N_{\rho}$
we write: 

\begin{equation}\label{Mxx}
\psi(\rho,\thet) = \sum_i\,F_i(\rho)\,\phi_i(\thet;\rho_n) 
\end{equation}
where $\phi_i(\thet;\rho_n)$ are eigenstates of the Hamiltonian at fixed $\rho_n$ distances.
Their calculation requires the solution of a bound state problem for a potential showing Coulomb 
singularities at $\theta=0$ and $\theta=\theta_{\mu}$. Treatment of Coulomb singularities requires 
the use of specific methods such as increasing the density of grid points, or alternatively the 
oscillation frequency of the basis functions, 
near the singularities. Mapped Fourier \cite{Fatt:96, Lemo:00}, 
Lagrange \cite{Baye:99} or Schwartz \cite{Schw:85,Duns:02}
 interpolations are examples of such methods which have been used for
cases with one singularity. 

In our case we have two singularities to deal with at fixed hyperradius. Therefore we have
used the
 auxilliary coordinate $x=2\,{\theta}/{\theta_{\mu}}-1$. 
This new variable is equivalent to one of the two hyperspherical elliptic angles 
\cite{Tols:95}, but constrained to be in the range [-1,1] with the
singularities being at the boundaries. 
If the  $\phi_i(\thet;\rho_n)$ functions are renormalized  as: 
$\phi_i(\thet;\rho)=(1-x^2)^{\frac{1}{2}}\bar{\phi_i}(x;\rho)$, 
both singularities of the potential are regularized.
The $\bar{\phi_i}(x;\rho)$ functions obey the differential equation:

\begin{eqnarray}
\left( -\frac{2\,\hbar^2}{m\,{\theta_{\mu}}^2\,\rho^2}\, \hat D + (1-x^2) (V - \frac{\hbar^2}{8\,m\,\rho^2})  \right) \bar{\phi_i}(x;\rho) 
\nonumber\\ \label{bphi}
= (1 - x^2) \epsilon_i(\rho) \bar{\phi_i}(x;\rho)
\end{eqnarray}
where 
\begin{equation} \hat D=(1-x^2)\frac{\partial^2}{\partial x^2} -2 x \frac{\partial}{\partial x} - \frac{1}{1-x^2}
\end{equation}

Eq.~\ref{bphi}  is solved by expanding 
 $\bar{\phi_i}(x;\rho)$ functions on the basis set of  the eigenvectors of $\hat D$,
which are the associated Legendre functions $P^1_n(x)$.  This yields
the generalized eigenvalue problem :

\begin{eqnarray}
\left( -\frac{2\,\hbar^2}{m\,\theta_\mu^2\,\rho^2} \,\bm{D} + \bm{W} \right) \,
\bar{\bm{\phi}}_i(\rho) \nonumber\\ = \epsilon_i(\rho) \, \bm{ O }\,  \bar{\bm{\phi}}_i(\rho)
\label{mphi}
\end{eqnarray}
where $\bar{\bm{\phi}}_i (\rho)$ is the vector of the unknown 
coefficients of the function $\bar{\phi_i}(x;\rho)$ in the Legendre basis set. 
The matrix representation  of the differential operator $\bm{D}$ is diagonal,
 with diagonal elements $-n(n+1)$.  $\bm{W}$ is the potential coupling
matrix and  $\bm{O }$ is the representation of $(1-x^2)$ in the Legendre
basis.
These matrices are calculated from a transformation of their diagonal 
representation in a grid of Gauss-Legendre quadrature points \cite{Core:92}. 
The system (\ref{mphi}) is transformed to a standard eigenvalue problem after 
left multiplication by $\bm{ O}^{-\frac{1}{2}} $.
%
%
The method is expected to converge fast with the size of the 
basis set (or equivalently with the number of grid points) since
 the use of a Gauss-Legendre scheme provides an adequately high density 
of points near  $x= \pm 1$. Since the wavefunctions concentrate 
in the vicinity of $x= \pm 1$ as energy decreases or as $\rho$ increases, 
we expect convergence to be most difficult for lower levels and large $\rho$. 
At large $\rho$, we also expect convergence to be more difficult to achieve for
 $O\mu$ states which are more compact than $p\mu$ states. We have computed the 
relative error for a given number $n_L$ of Legendre basis functions, using 
$n_L=450$ as a reference. For $n_L=150$, we obtained 86 states converged 
better than $10^{-8}$ of relative error up to $\rho\sim a_0$. 
For larger $\rho$ values, the convergence starts to deteriorate, first for the 
ground state $O\mu(n=1)$, then for the first excited state $n=2$ 
near $\rho=2\,a_0$...Using larger values of $n_L$ delays the 
limit of convergence deterioration. For instance,
 with $n_L=250$, the ground state is converged with $10^{-8}$ 
relative precision up to $\rho\sim 5\,a_0$, up to $\rho\sim 10\,a_0$ with $n_L=350$. 
This latter value of $n_L$ was used 
to generate the results shown below, and provided a basis 
converged better than $10^{-4}$ up to $\rho\sim 40\,a_0$ 
for all states except the first four ones which do not play any 
significant role in the charge transfer process considered here. 
As expected, among all these states, the ones corresponding to 
bound states of $p\mu$ are the easiest to converge : the relative error 
at $\rho=40\,a_0$ for $p\mu$(n=1) is $10^{-10}$ with 
$n_L=250$ and $10^{-12}$ with $n_L=350$. High accuracy 
on the $p\mu(n=1)$ channel is especially important for our 
present study since we consider initial relative kinetic energies 
 as low as $10^{-6}$ eV.    

The close-coupling equations are integrated
along the hyperradius $\rho$ using the de Vogelaere algorithm \cite{Lepe:86}. 
In order to have convergence of the transfer probabilities to better than 1 \% 
in the energy range considered here, we included 29 channels:
($(p\,\mu)_{n=1-4} + \mO$ and
$p + (\mu\,\mO)_{n=1-25}$).  The integration of the coupled equations
was performed from the origin to
 $\rho_{\mbox{end}}\sim 30\,a_0$.

The asymptotic analysis has been performed using
the appropriate Jacobi coordinates 
for the entrance  and   for the product channels.
The asymptotic states were expressed  as products of Coulombic bound states
 and translational functions.  For the product channels,
the translational functions were approximated by simple plane waves. However,
since the initial relative translational energy can be very small, the entrance channel
has to be described with care to account for residual long range interactions.
For the pure coulombic potential, this residual interaction has
a $1/R^2$ dependence (charge-dipole interaction).
 In this case, Bessel functions of imaginary order have been used as a basis
set in $R$. For the screened Thomas-Fermi potential, wave functions depending on $R$
have been numerically computed by inward integration from very large distances.
These asymptotic wave functions and their derivatives are projected
onto the hyperspherical basis functions $\phi_i(x;\rho_{\mbox{end}})$ \cite{Lepe:86}.
Transfer probabilities are then obtained by a standard S-matrix analysis.
 \section{Results of the calculations}
  
\begin{figure}
\includegraphics[width=8.cm]{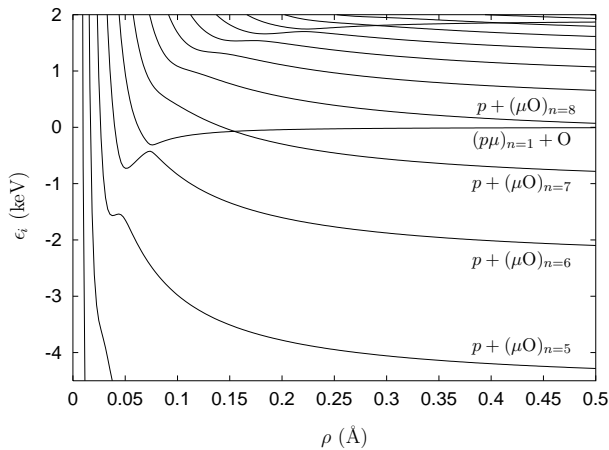}
\caption{\label{figure_a1_2}{Adiabatic energies as a function of the
hyperradius.}}
\end{figure}

We have performed calculations for collision energies in the range $10^{-6}-10^{3}$ eV.
In Fig.~\ref{figure_a1_2} we present the adiabatic energies $\epsilon_i$ (see Eq.~\ref{bphi})  as a function of the
hyperradius $\rho$.  The origin of energies has been chosen
to coincide with the asymptotic limit of the entrance channel
$(p\,\mu)_{n=1} + \mO$. Thus the calculations cover the
energy range between this limit and the $p + (\mu\,\mO)_{n=10}$
threshold.

The dynamics of muon transfer can be qualitatively understood by inspection
of this figure. Starting in channel $(p\,\mu)_{n=1} + \mO$,
the system crosses diabatically the channel 
 $p + (\mu\,\mO)_{n=7}$. Muon transfer is completely negligible 
as the coupling is very small compared with the collision energy.
The couplings  to  channels $p + (\mu\,\mO)_{n=6}$ and $p + (\mu\,\mO)_{n=5}$ are 
larger as evidenced by  avoided
crossings.  The other
channels $p + (\mu\,\mO)_{n<5}$ are weakly coupled to the initial one
and they are not expected to be populated significantly.  

\begin{figure}
\includegraphics[width=8.cm]{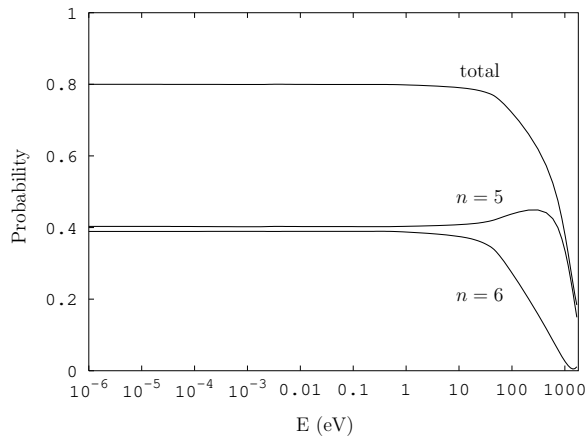}
\caption{\label{figure_a1_3}{Partial transfer probabilities calculated with the
pure coulombic potential (bare oxygen nucleus).}}
\end{figure}

\begin{figure}
\includegraphics[width=8.cm]{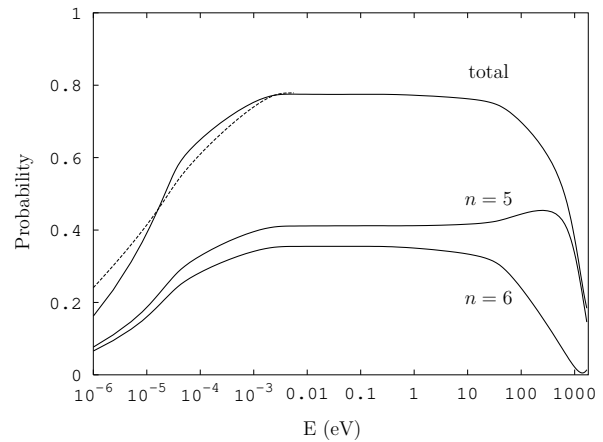}
\caption{\label{figure_a1_4}{Multichannel calculated partial and total transfer 
probabilities for the Thomas-Fermi potential.
The dashed curve is the result of the 
 Eq.~(\ref{sim})}}
\end{figure}

In Figs.~\ref{figure_a1_3} and \ref{figure_a1_4}
we present the muon transfer probabilities into the
different product channels $p + (\mu\,\mO)_{n=5,6}$ together with the total
transfer probability, for the pure coulombic (C) and the Thomas-Fermi (TF) potentials, respectively. The two channels are
about equally populated in both cases and the total transfer probability
is high (80 \%) for energies in the  intermediate range (between $10^{-2}$ and $10^2$ eV).
This can be understood in terms of non-adiabatic transitions
in a simple 3 channels model ( $(p\,\mu)_{n=1} + \mO$ and $p + (\mu\,\mO)_{n=6,5}$).
Using the Landau-Zener expressions for the two crossings,
 a total transfer probability of $\sim 0.6$ between 0 and $10^2$ eV  and
a slow decrease at higher energies is obtained. This is in qualitative agreement with the fully
converged results.

The main difference between the results for
the C and TF potentials is  at low energies. Whereas for the C potential
the probabilities are constant at low energies, for the TF potential there is
a typical threshold behavior. This is what is expected for short range potentials \cite{Land:77}.
Indeed, in the colinear configuration the C potential behaves asymptotically as $1/R^2$ 
(charge-dipole interaction) while the TF potential goes to zero much faster.
At low energies,  the amplitude of the wave function for $1/R^2$ potentials
near the origin is constant, while for short range potentials
it decreases as energy goes to zero \cite{Land:77}. Thus the low energy  dependence 
of the transfer probabilities is enterily
determined by the asymptotic behavior of the potentials. This suggests that the dynamics
can be described by a two step mechanism: a transmission in the incident channel through the long range
part of the potential, followed by non adiabatic transitions in
the interaction region. The charge transfer probability for low energies will be then  given by
\begin{equation}\label{sim}
P(E) = \vert T(E)\vert^2\,P_{max}
\end{equation}
where $T(E)$ is the transmittance in the incoming channel and $P_{max}$ the
plateau
transfer probability.

The transmittance $T(E)$ can be estimated by a simple 
one channel calculation with an effective potential
of the form
\begin{equation}\label{eff}
V_{\mbox{eff}}(R) = -\frac{\alpha\,Z_{\smO}^\ast(R)}{R^2};\quad \alpha=\frac{3\,\hbar^2}{2\,m_{p,\mu}}
\end{equation}
for $R>R_0$ and $V_{\mbox{eff}}(R) =V_{\mbox{eff}}(R_0) $ for $R<R_0$.
The transfer radius has been taken at the avoided crossing with the
 $p + (\mu\,\mO)_{n=6}$ channel. 
 For $R<R_0$ the solution of this one-channel problem
is given by
\begin{equation}
F(R) = T(E)\,\frac{\sin(K\,R)}{K^{1/2}}
\end{equation}
with $K=(2m_{\smO,p\mu}(E - V_{\mbox{eff}}(R_0)))^{1/2}/\hbar$,
while for $R\to\infty$ we have $F(R) =\sin(k\,R+\delta)/k^{1/2}$ with $k=(2m_{\smO,p\mu}E)^{1/2}/\hbar$.
Numerical solution of the one channel problem thus provides directly $T(E)$. For the
pure Coulombic potential ($Z_{\smO}^\ast=8$) we found $T(E)=1$, consistent with the
result of our multichannel calculations which gives a constant transfer probability
at threshold (see Fig.~\ref{figure_a1_3}). This unusual behavior is due to the
particular $1/R^2$ dependence of the effective potential \cite{Land:77}. The shorter
range TF potential gives a more standard threshold behavior, with a transmittance
going to 0 as energy decreases. This is apparent in  Fig.~\ref{figure_a1_4} where  
the total transfer probabilities for the TF potential as a function of the energy
are presented for
both, the multichannel calculation  and 
the simple estimation using Eq.~(\ref{sim}). The two calculations agree within 10\%.

\begin{figure}
\includegraphics[width=8.cm]{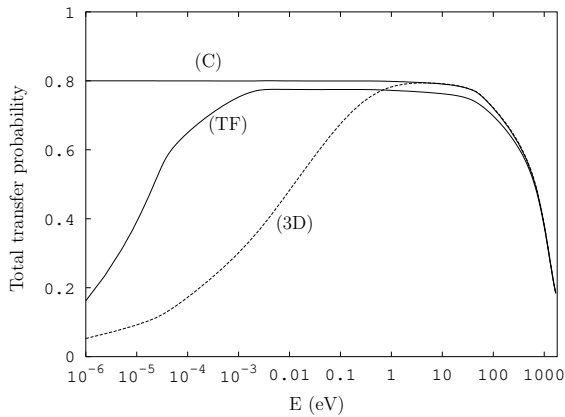}
\caption{\label{figure_a1_5}{Total transfer probabilities for the colinear model 
((C): pure Coulombic potential, (TF): Thomas-Fermi model) and 3D estimate
obtained from Eq.~\ref{sim}.}}
\end{figure}

This  suggests a simple procedure for 
estimating the energy dependence in 3 dimensions (3D). In 3D,
the ion-dipole interaction perturbes the $p\mu_{n=1}$ initial state at
second order through a $-1/R^4$ term
 \cite{Gers:63}.
So even for the pure coulombic case there will be a threshold behavior
 at low energies. We have thus performed a
one channel calculation of the transmittance $T(E)$ with the 3D functional form of the  potential. 
This $T(E)$ was used in Eq.~(\ref{sim}) together with $P_{max}$ provided
by the colinear Coulombic potential to yield an approximate 3-dimensional
transfer probability.
Since the colinear configuration is the most favorable
configuration, this procedure provides an upper limit of the total transfer probability.
The results are presented in
figure \ref{figure_a1_5} together with those of the colinear calculations.
Clearly, the threshold is displaced towards higher energies ($10^{-1}$ eV).
Thus we expect that for thermal energy collisions ($4\,10^{-2}$ eV) the
muon transfer probability will be of the order of 0.6 or less and will increase
smoothly with energy in the range $10^{-2}-10^{-1}$ eV.

\section{Discussion}

In order to make contact with the experiments of
Werthm\"uller \etal \cite{Wert:98}, we have calculated the
muon transfer rate according to:
\begin{equation}\label{rate}
\lambda(E) = N\,v\,\sigma(E)
\end{equation}
with $N$ being the number density of liquid hydrogen ($4.25\,10^{22}$ cm$^{-3}$), $v = \sqrt{2\,E/m_{\smO,p\mu}}$  the relative velocity
and
\begin{equation}\label{sigma}
\sigma(E) = \frac{\pi}{k^2}\,P(E)
\end{equation}
 the muon transfer cross section with $k=\sqrt{2\,m_{\smO,p\mu}\,E}/\hbar$.
In writing Eq.~(\ref{sigma}), we have assumed only one partial wave
with zero  angular momentum.  
This is valid for collision energies
up to  0.1--0.2 eV, corresponding to the height of the
barrier in the entrance channel for the $p$ partial wave. 

\begin{figure}
\includegraphics[width=8.cm]{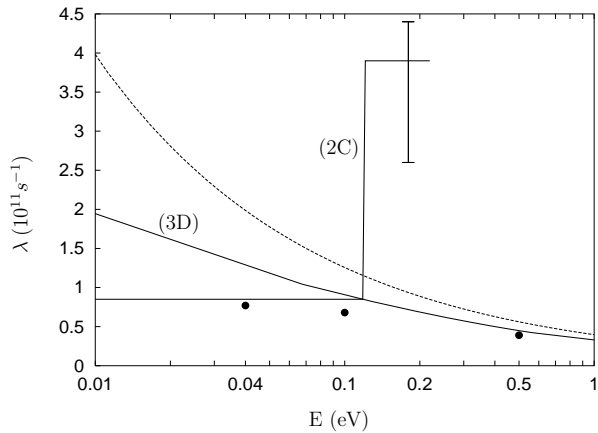}
\caption{\label{figure_a1_6}{Estimated transfer rates obtained in this work (3D). 
The dashed curve represents the rate corresponding to a maximum
transfer probability $P(E)$ of unity.
The points are the calculated values of Sultanov and Adhikari \cite{Sult:00}
using the Fadeev formalism.
The solid straight lines (2C) are the rates determined from fitting the experimental data of
Ref.~\cite{Wert:98} with the two-component model.}}
\end{figure}

In Fig.~\ref{figure_a1_6} we present our 3D estimates   together with 
the results of the calculations of 
Sultanov and Adhikari \cite{Sult:00} based on a two-states
approximation of the Fadeev-Hahn equations. The agreement 
is good, our results
being slightly higher than theirs. This was expected since our 3D
estimation is based on the 
colinear calculations which provides an upper limit
to the transfer probabilities. In addition, they have consider only the
O$\mu_{n=5}$ final channel 
 From Eqs.~(\ref{rate}) and (\ref{sigma}) it follows that for a constant transfer probability $P(E)=1$ the rate
should decrease as $1/\sqrt{E}$. This is also plotted in Fig.~\ref{figure_a1_6}. We note
that our calculated rates decrease with energy slower than those for a constant 
transfer probability. The reason is that our calculated probabilities actually increase
(although slower than $1/\sqrt{E}$) in the energy range considerde here as can be
seen in Fig.~\ref{figure_a1_5}.

In Fig.~\ref{figure_a1_6} we have also represented the
rates of the two-component model used in Ref.~\cite{Wert:98}  to fit the experimentally
observed X-ray emission from excited muonic
oxygen. While for thermal energies  there is an order of magnitude
agreement between our estimated values and the two-component  model,
there is a clear disagreement for epithermal energies. The two-component
model predicts an increase of the rates going from thermal
to epithermal energies in contradiction with our predictions and those
of Sultanov and Adhikari \cite{Sult:00}.
Moreover, for energies above 0.1 eV
the rate predicted by the two-component model
is three times larger than the maximum values of the rate
 obtained by assuming a muon transfer probability per collision of unity (see Fig.~\ref{figure_a1_6}).

\section{Conclusions}

We have presented colinear calculations of muon transfer rates 
between muonic hydrogen and oxygen for relative translational  energies between
$10^{-6}$ and $10^3$ eV. For the lower energies (below $10^{-1}$ eV)  the de Broglie wavelength is much larger
than the characteristic distance of the potential interaction and
the problem corresponds
to an ultra-cold collision.
A simple procedure to estimate the energy dependence
of the muon transfer rate in three dimensions was proposed.

Our results show that the muon transfer rate decrease going from thermal
to epithermal energies, in agreement with
previous theoretical calculations \cite{Sult:00} but in contradiction
with the assumptions of a two-component model
used  in Ref.~\cite{Wert:98} to interpret the experimental data on this system.

In the experiment the X-ray emission shows a bi-exponential
behavior. In the two-component model of Werthm\"uller \etal \cite{Wert:98}, it was 
assumed that the prompt emission is  due to muon transfer from
epithermal muonic hydrogen while the delayed emission,
 similar to what is observed when oxygen is replaced by other atoms,
is associated to thermalized muonic hydrogen.
Assuming that the relative amounts of epithermal and thermal
species are equal, Werthm\"uller \etal concluded that the rate of muon transfer from epithermal muonic hydrogen
to oxygen should be larger than the one from thermalized ones by a factor of almost 4.
They suggest that a resonance could be responsible for this increase.
We have shown that for s-waves it is not possible to obtain such
an increase in the transfer rate, even if assuming a transfer probability per
collision of one. Thus, if resonance effects exist they have to
be due to the contribution of higher order partial waves.
In an elastic cross section calculation, Kravtsov \etal \cite{Krav:96}
have shown that for muonic hydrogen colliding 
with an oxygen nuclei, the contributions of $p$ and $d$ waves are
small as compared to the $s$ wave in the range
0--0.2 eV and that only at $\sim$1.5 eV there is narrow
resonance involving the $d$ wave. This will thus imply that
the epithermal muonic hydrogen has much higher translational
energy than what has been assumed in Ref.~\cite{Wert:98}.
Anyway, this conclusion is based on a calculation of the
elastic scattering. It could be of interest to check whether
it is valid also for the muon transfer process as well.
%
%

In this work the calculations were performed in a restricted colinear
configuration. This approximation is known to give poor quantitative
results in many cases. However for the system considered here,
the colinear configuration is the most favorable for the transfer process. Also,
since  the colinear configuration provides
the minimum energy path for the reactance channel, even for an initial non colinear configuration
there will
be efficient orientational effects  in particular for low translational energies. In
addition we have introduce a correction to the low energy dependence of the rate
in order to take into account the correct asymptotic behavior of the
potential in 3-dimensions. Therefore, we believe that although 
quantitatively the results may change if a full 3-dimensional
calculation is performed, the conclusions, in particular 
concerning the energy dependence of the muon transfer rates,
will remain valid.
A full 3-dimensional treatment is not impossible however, although
quite computationnally demanding. It will definitetly be very important
to further improve our detailed understanding of this 
reaction.
It would also be most interesting to study the energy dependence
of the muon-transfer rates
for other elements, such as C or S, for which the rates
seem to behave differently as a function of the energy.
This work is in progress.

\section{Acknowledgments}

We thank J.M. Launay for helpful discussions, the IDRIS
for an allocation of CPU time on a NEC SX5 vector processor, and NATO for a collaborative linkage grant 
PST.CLG.978454 between Bulgaria and France.   

\end{document}